# Fate of global superconductivity in arrays of long SNS junctions

Serena Eley[1], Sarang Gopalakrishnan[1], Paul M. Goldbart[2], Nadya Mason[1]

**Abstract**

Normal-metal films overlaid with arrays of superconducting islands undergo Berezinskii-Kosterlitz-Thouless (BKT) superconducting transitions at a temperature $T_{BKT}$. We present measurements of $T_{BKT}$ for arrays of mesoscopic Nb islands patterned on Au films for a range of island spacings $d$. We show that $T_{BKT} \sim 1/d^2$, and explain this dependence in terms of the quasiclassical prediction that the Thouless energy, rather than the superconducting gap, governs the inter-island coupling. We also find two deviations from the quasiclassical theory: (i) $T_{BKT}$ is sensitive to island height, because the islands are mesoscopic; and (ii) for widely spaced islands the transition appears to lead, not to a superconducting state, but to a finite-resistance "metallic" one.

Arrays of superconducting islands placed on thin normal-metal films offer tunable realizations of two-dimensional superconductivity [1–3]. In such superconductor—normal-metal—superconductor (SNS) arrays, the superconducting islands are coupled to one another through the proximity effect (i.e., the diffusion of Cooper pairs from the superconducting islands into the normal metal), which can give rise to global phase coherence through a Berezinskii-Kosterlitz-Thouless (BKT) transition below a temperature $T_{BKT}$ [1,2]. The dependence of $T_{BKT}$ on island spacing can elucidate the physical processes and energy scales governing the inter-island coupling; however, this dependence has not previously been measured systematically. The standard theoretical analysis of SNS arrays, due to Lobb, Abraham, and Tinkham (LAT) [2], predicts that $T_{BKT}$ should depend on island spacing $d$ according to the relation $T_{BKT} \approx \Delta \exp(-d/\xi_N(T_{BKT}))$, where $\Delta$ is the superconducting gap on the islands, $\xi_N(T) = \sqrt{\hbar D/k_B T}$ is the normal-metal coherence length, and $D$ is the normal-metal diffusion constant. The LAT analysis is based on the Ginzburg-Landau theory, which is valid only when $\Delta$ is smaller than any other relevant energy scale. However, for sufficiently widely spaced islands, $\Delta$ is much larger than the Thouless energy $E_{Th} = \hbar D/d^2$ (which is inversely proportional to the time for the diffusion of electrons between the islands). In this regime, one can adapt the quasiclassical theory of Refs. [4,5] to show that the proximity

---

[1] Department of Physics and Frederick Seitz Materials Research Laboratory, University of Illinois Urbana-Champaign, Urbana, IL 61801, USA
[2] School of Physics, Georgia Institute of Technology, 837 State Street, Atlanta, GA 30332, USA

coupling between islands depends *only* on $E_{Th}$ and not on $\Delta$ [4–7]; as we argue below, the quasiclassical theory then implies that $T_{BKT} \sim 1/d^2$. It has also been predicted [8–10] that for arrays having sufficiently large $d$, the system undergoes a quantum phase transition to a zero-temperature metallic state, rather than a superconducting one. Such a metallic state would necessarily be unconventional, as Anderson localization precludes conventional zero-temperature metals in two-dimensional systems [11,12].

To test these predictions, we have systematically studied the dependence of the superconducting transition on island spacing for arrays of widely spaced islands (i.e., in the regime where $E_{Th} \ll \Delta$). In a recent paper [13], we showed that the spacing dependence of $T_{BKT}$ deviated strongly from the LAT prediction. In this Letter, we measure $T_{BKT}$ over a wider range of island spacings, and find that it depends on spacing as $1/d^2$, which is consistent with the $\Delta$-independent behavior predicted by the quasiclassical theory. We also present measurements of the temperature dependence of the critical current $I_c(T)$, and find that the behavior of $I_c(T)$ in arrays is similar to that of single SNS junctions and SQUIDs [4,6,14]; however, such measurements of $I_c$, while consistent with the quasiclassical predictions, only indirectly address the role of the Thouless energy. We argue that the dependence of $T_{BKT}$ on island spacing in SNS arrays exposes the role of the Thouless energy more directly: because the quasiclassical theory implies that the only relevant energy scales are $E_{Th}$ and $k_B T$, it follows from dimensional analysis that $T_{BKT}$ must be proportional to $E_{Th}$, and thus that $T_{BKT} \sim 1/d^2$. We also show that our results deviate from the quasiclassical predictions in two ways. First, we find that $T_{BKT}$ depends nontrivially on the height of the islands; this reflects the mesoscopic character of the islands, which distinguishes our experiments from those of [1]. The second discrepancy we find is that, for the widest spaced islands we measured, the superconducting transition appears to be interrupted, and the resistance flattens out at a nonzero value. This flattening indicates that the system might be metallic at $T = 0$.

The devices studied in this Letter consist of triangular arrays of Nb islands (see Figure 1) fabricated on 10-nm thick Au films, which are patterned for four-point measurements, on Si/SiO$_2$ substrates [13]. Each substrate contains up to six film/array devices, which are identical except for their systematically varied island spacings. The diameter of each Nb island is 260 nm; the island height (i.e., Nb thickness) is identical for all devices on a single substrate, but ranges from 87 nm to 145 nm for the sets of devices presented in this manuscript. The islands are composed of columnar grains ~ 30 nm in diameter [13], and have a dirty-limit coherence length $\xi_0^{Nb} \approx 27\ nm$, for an approximate mean free path $\ell \approx 8\ nm$. The edge-to-edge island spacings $d$ range from 90 nm to 1.24 $\mu m$, and the number of islands per array varies from 11,400 to 155,800. The Au film has an estimated diffusion constant $D \approx 95\ cm^2/s$ for a mean free path of $\ell \approx 13\ nm$ and a temperature-dependent coherence length $\xi_N(T) \approx 270\ nm/\sqrt{T}$, where $T$ is in units of K. Resistance measurements (through the Au film) were performed using standard low-frequency, ac lock-in techniques in either a pumped He-4 cryostat, a He-3 cryostat, or a He-3/He-4 dilution refrigerator. Figure 1 shows the temperature-dependent normalized resistance for arrays having islands spaced 190 to 340 nm apart. The islands become superconducting at a transition temperature $T_1$, and superconductivity across the array appears below a critical temperature $T_{BKT} < T_1$ [13].

The upper inset to Figure 1 shows the current-biased differential resistance $dV/dI$ for an array having $d = 240\ nm$ at $T = 2.4\ K$ (i.e., below $T_{BKT}$); the shape of the curve is typical of results observed in all of our arrays. Differential resistance was determined by differentiating the $IV$ characteristics. To minimize the effects of Joule heating, these $IV$ characteristics were measured using current pulses, with a current-on time of 3.5 $ms$ and -off time of 3 $ms$. Peaks are evident at positive and negative currents; the lower-current (inner) peaks mark the critical current $I_c$ across the proximity-coupled Au film (i.e., where the $IV$ curves become approximately Ohmic) [15]. The higher-current (outer) peaks correspond to the critical current of the islands. In this Letter, we focus on the inner peaks, or the $I_c$ corresponding to the BKT transition of the film. Figure 2a shows these inner peaks for devices having various island spacings. As can be seen in the Figure, the peaks move toward lower bias—i.e., $I_c$ decreases—for larger island spacings. In Figure 2b, we show $I_c$ as a function of temperature $T$ for four different arrays, where $I_c$ was extracted from the peaks in $dV/dI$. The solid curves are fits to the temperature-dependent, quasiclassical equations for a single, diffusive SNS junction, namely $I_c(T) = a\frac{E_{Th}}{eR_N}(1 - be^{-aE_{Th}/3.2k_BT})$, where $R_N$ is the normal resistance, and dimensionless parameters $a$ and $b$ depend on $\Delta/E_{TH}$ [4]. We treat $R_N$ as a fitting parameter, $R_N^{Fit}$ noted in the table in Figure 2, which deviates somewhat from the actual $R_N$, as has been previously observed in single SNS junctions [4,6,16–18]. As can be seen in Figure 2b, the quasiclassical predictions fit the data well, and the fit values (noted in the table in Figure 2) are close to those predicted for very widely spaced single junctions ($\Delta \gg E_{Th}$), namely $a = 10.82$ and $b = 1.30$, which suggests that the SNS arrays display behavior similar to that of single junctions.

We now discuss the dependence of $T_{BKT}$ on island spacing. The value of $T_{BKT}$ was extracted from the temperature dependence of the $IV$ curves (see Figure 3a); these curves exhibit behavior consistent with a BKT transition, which is a hallmark of 2D superconducting systems. For $T < T_{BKT}$, the $IV$ characteristics are nonlinear such that $V \propto I^{\alpha(T)}$, where $\alpha(T) = 2(T_{BKT}/T) + 1$ for $T \leq T_{BKT}$ and $\alpha(T) = 1$ for $T > T_{BKT}$; at the transition $\alpha(T_{BKT}) = 3$ [19]. Following standard practice [18], we mark the temperature of this "Nelson-Kosterlitz" jump in $\alpha(T)$ as $T_{BKT}$, as shown in Figure 3b. In Figure 3c we plot the spacing dependence of $T_{BKT}$, which is well captured by the form $T_{BKT} \sim 1/d^2$, especially for the arrays having the most widely spaced islands. The suppression of $T_{BKT}$ with increasing island spacing is clearly more rapid than that predicted by the LAT theory, as is evident in the main panel of Figure 3c.

The observed trend of $T_{BKT} \sim 1/d^2$ implies that $k_B T_{BKT}/E_{Th}(d) =$ constant. The data show that the constant is of order unity. This linear relationship between $T_{BKT}$ and $E_{Th}$ follows from the quasiclassical prediction [21] that $E_{Th}$ is the unique energy scale governing the properties of a long SNS junction. One can understand this prediction in the following heuristic terms [21,22]. The proximity effect is a consequence of Andreev reflection [21], whereby normal-metal electrons with energy $\varepsilon < \Delta$ above the Fermi energy incident at the NS interface are retroreflected as holes; at the interface, the phase between the electron and retroreflected hole is set by the superconductor. As long as the electron and hole stay phase-coherent with one another while propagating through the normal metal, they carry information about the superconducting phase and thus mediate the proximity effect. It can be shown [23] that, at zero temperature in a diffusive metal, an electron with energy $\varepsilon$ above the Fermi surface dephases with its retroreflected hole after traveling a distance of order $L_0 = \sqrt{\hbar D/\varepsilon}$, where $D$ is the diffusion constant in the normal metal; hence, only electrons with energy $\varepsilon \lesssim \hbar D/d^2 = E_{Th}$ carry superconducting phase information and can be involved in the proximity effect. (This analysis is valid in the "long-junction" limit, where $E_{Th} < \Delta$.) Thus, the Thouless energy is the unique energy scale determining the zero temperature properties of SNS arrays in the long-junction limit, and the spacing-dependence of observables can be established by dimensional analysis (so that, e.g., $I_c R_N \sim E_{Th}$ [4]). At finite temperatures, thermal dephasing adds an additional energy scale $k_B T$. Therefore, by dimensional analysis, all finite-temperature properties are determined by the ratio $E_{Th}$ to $k_B T$ (which can be equivalently rearranged as the ratio of $\xi_N$ to $d$). In particular, this holds true for the transition temperature, so that $T_{BKT} \sim E_{Th}/k_B \sim 1/d^2$, which is indeed the trend we observe experimentally.

The quasiclassical theory not only predicts that $T_{BKT} \sim 1/d^2$, but also implies that the constant of proportionality should be universal for islands having dimensions much larger than the superconducting coherence length $\xi_0^{Nb}$ (since no scales aside from $E_{Th}$ are relevant). However, we find that the ratio of $T_{BKT}$ to $E_{Th}$ is not universal but depends on the island height, as can be seen in the inset of Figure 3c. This dependence is likely a signature of the mesoscopic, granular character of the islands [13], as $\xi_0^{Nb} \approx 27 \, nm$ is comparable to the grain size within the islands (though smaller than the island size). Because the island height changes the prefactor but not the $1/d^2$ scaling of the transition temperature, we conjecture on dimensional grounds that

$$\frac{T_{BKT}}{E_{Th}} = f\left(\frac{G\delta}{\Delta}\right)$$

where $\delta$ is the spacing of energy levels in the grains that constitute each superconducting island, and $G$ is the dimensionless conductance of the normal metal; the product $G\delta$ is thus a measure of the inverse "dwell time" of an electron on one of the superconducting grains (see, e.g., Refs. [24–26]). For the grains discussed here, $\delta \sim 0.8 \, \mu eV$, and the conductance of the gold film, $G \sim 950$; thus, the ratio $G\delta/\Delta$ is of order unity, consistent with the appreciable height-dependence of $T_{BKT}$ that is observed experimentally. In principle, a second candidate for a relevant mesoscopic energy scale is the charging energy on each island. However, the transition temperature of an array of closely-spaced islands (e.g., for $d = 90 \, nm$ and $\xi_N(T \approx 8.9 \, K) \approx d$) is approximately equal to the critical temperature an un-patterned, Nb-Au bilayer film, $T_c \approx 8.9 \, K$. This leads us to believe that the islands are well coupled to the Au film, making it unlikely that charging effects are significant.

A second discrepancy between the quasiclassical theory and our experimental findings concerns the low-temperature behavior of arrays in which the island spacing is greater than ~ 700 nm. As can be seen in Figure 4, for these arrays the resistances do not vanish at the expected transition temperatures; instead, they first begin to drop and then flatten out to fixed, nonzero values. As the island spacing is increased beyond ~ 700 nm, the asymptotic low-temperature resistance increases until it eventually exceeds the normal-state resistance of the Au film. Both the LAT and the quasiclassical theories preclude such a flattening of the resistance, as they predict that once the superconducting transition begins, the inter-island coupling should continuously increase as the temperature is decreased. At present, we do not have a clear understanding of this low-temperature resistance, although artifacts of the experimental set-up can be excluded. In particular, we have determined that if the excitation current is greater than the critical current, the superconducting transition simply shifts to lower temperatures instead of being interrupted by a plateau. We have also excluded the possibility of insufficient sample cooling by determining that a temperature sensor mounted directly on a sample holder cools below $25 \, mK$. Moreover, Joule heating is unlikely, as we observe that higher-resistance samples (i.e., those with more widely spaced islands) seem to plateau at lower temperatures, an opposite trend from Joule heating. Thus, we are led to conjecture that, for sufficiently large island spacing, global phase coherence across the array is interrupted, and instead, a finite-resistance, "metallic" state is achieved. Similar resistance plateaus have previously been observed in single Josephson junctions [27] and in quantum wires [28], and have been attributed to macroscopic quantum tunneling of vortices. For the SNS arrays discussed here, the salience of the mesoscopic character of the islands for $T_{BKT}$, and the granularity of the islands, suggest that the phase-fluctuation-based mechanisms discussed in Refs. [8,9] may also apply. If so, then the finite-resistance state achieved at low temperatures should consist of well-separated superconducting puddles; we shall experimentally investigate this possibility in future work.

To summarize, we have presented systematic measurements of the superconducting transition temperature of SNS arrays as a function of island spacing, and thus directly established the key role played by the Thouless energy in determining inter-island phase coherence. While the importance of the Thouless energy is an expectation of the quasiclassical theory, our results deviate from quasiclassical expectations in some respects, notably in the dependence of the transition on island height and in the appearance of a low-temperature metallic state for very large island spacings; the origin and properties of this metallic state remain open questions for future investigation.

**Acknowledgement**

This research was supported by the US Department of Energy (DOE), Division of Materials Science (DMS) under grant DE-FG02-07ER46453 through the Frederick Seitz Materials Research Laboratory at the University of Illinois at Urbana-Champaign.

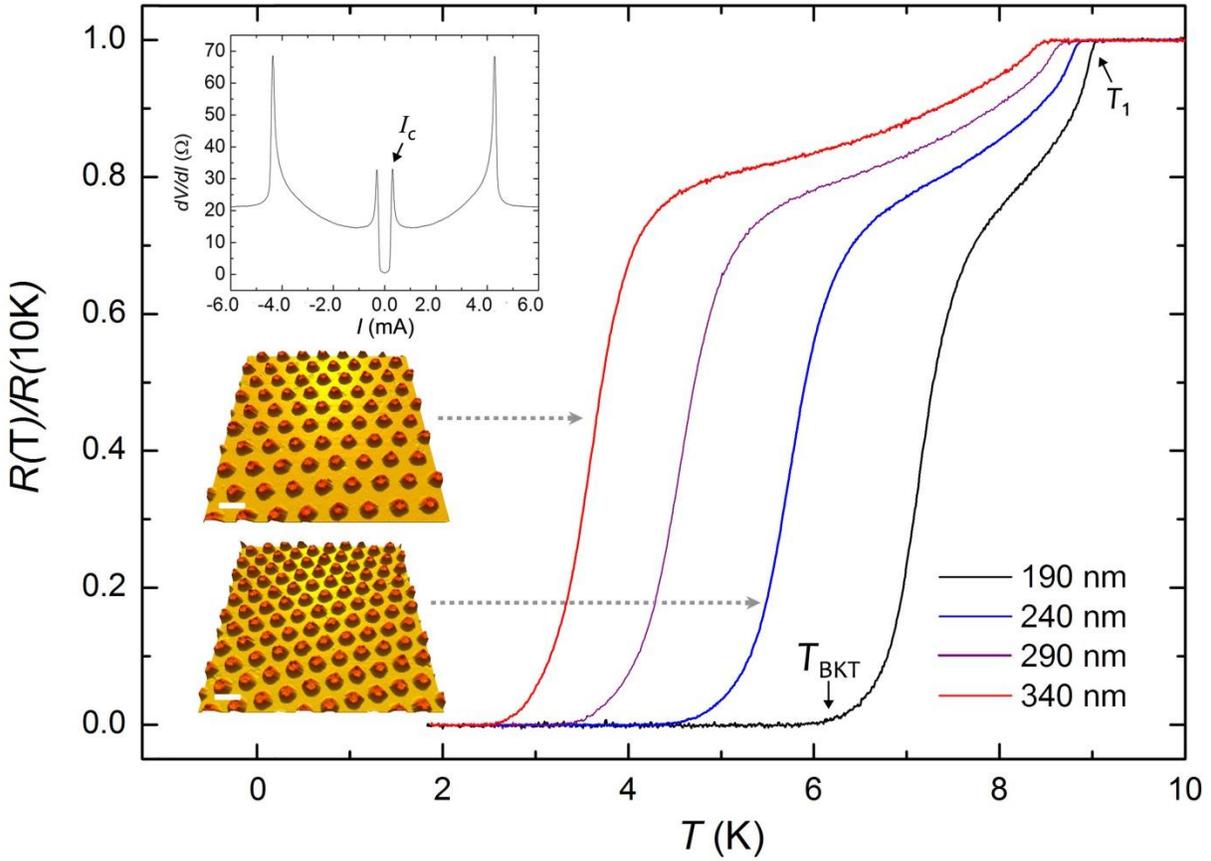

FIG. 1. Resistance $R$ (normalized to values at 10 K) vs. temperature $T$ for four arrays, each having different island spacings $d$. For the array having 190 nm-spaced islands, the transition temperature of the islands is marked as $T_1$ and the BKT transition temperature is marked as $T_{BKT}$. Lower inset: AFM images of Nb islands (red) on Au films (yellow), having 340 nm- and 240 nm-spaced islands (scale bar is 500 nm). Upper inset: typical differential resistance $dV/dI$ characteristics for $T < T_{BKT}$; here, $d = 190$ nm and $T = 2.4$ K. The inner peaks, denoted $I_c$, correspond to the critical current of the proximity-coupled film; the outer peaks correspond to the critical current of the islands.

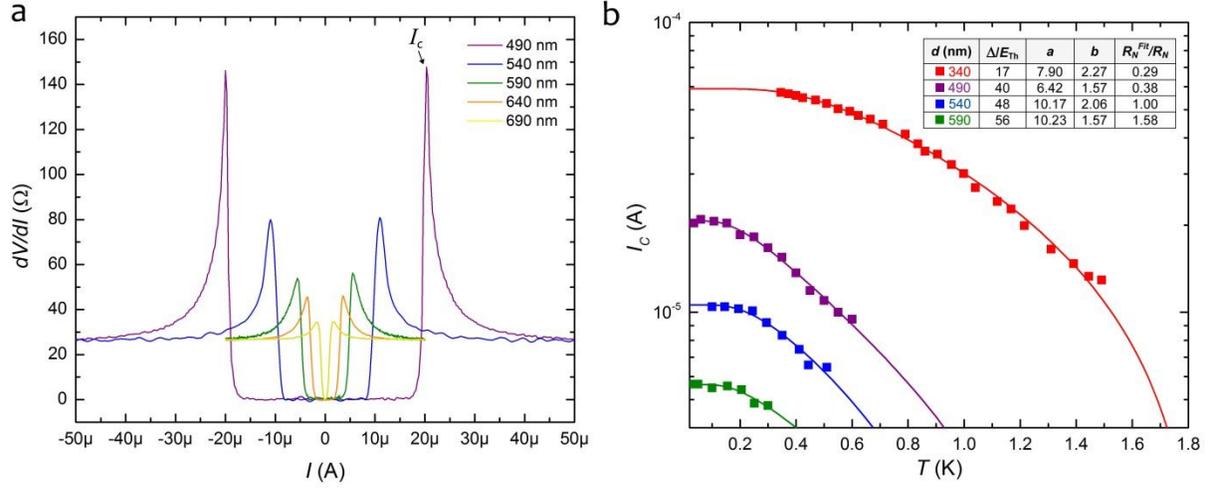

FIG. 2. (a) Differential resistance vs. bias current as a function of island spacing, at $T \approx 30$ mK. The critical current $I_c$ is extracted from the peak position. (b) $I_c$ vs. $T$, for four island spacings. Solid lines are fits to the quasiclassical theory, as discussed in the text; parameters relevant to the fits are given in the table.

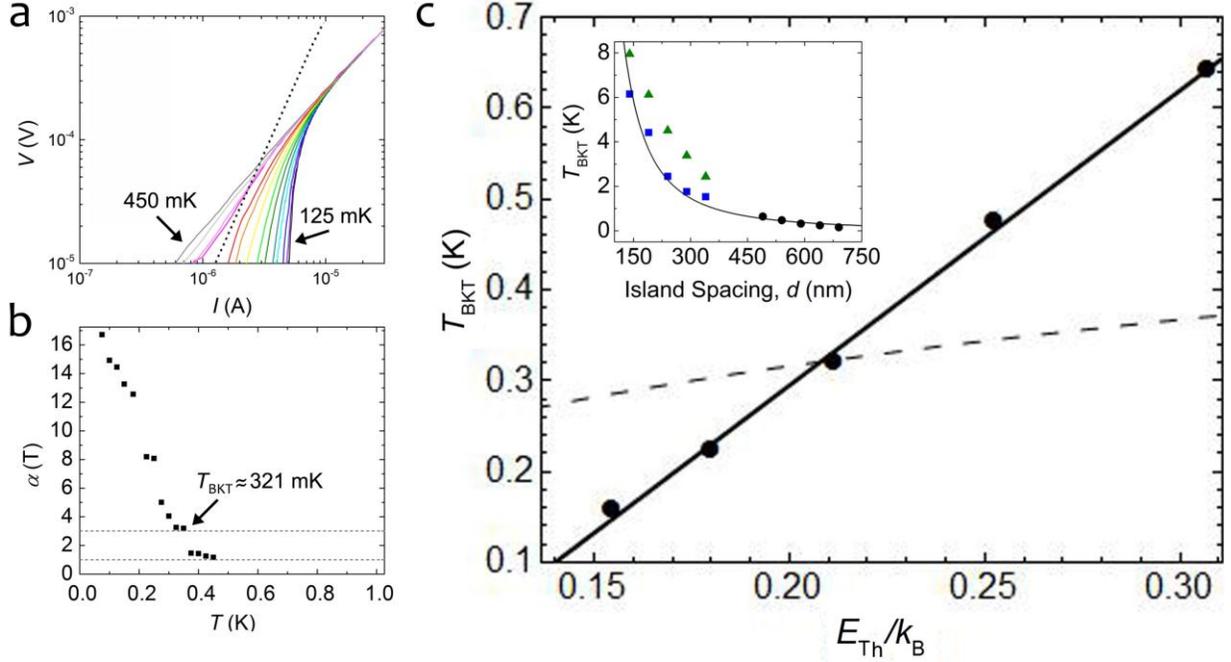

FIG. 3. (a) Logarithmic current-voltage (*IV*) characteristics for 590-nm-spaced islands, at temperatures near $T_{BKT}$, ranging from 125 mK to 450 mK at intervals of 25 mK; dashed line separates Ohmic behavior above $T_{BKT}$ from non-Ohmic ($V \sim I^{\alpha(T)}$) behavior below $T_{BKT}$. (b) The exponent $\alpha$ as a function of temperature for the same array. $T_{BKT}$ is taken to be the temperature at which $\alpha = 3$ [19]. (c) Main figure: $T_{BKT}$ vs. the Thouless energy $E_{Th} \sim 1/d^2$ for widely spaced islands ($d > 450$ nm). Solid black line shows linear fit. Dashed black line shows the fit to the LAT theory. Inset: $T_{BKT}$ vs. $d$ over a wider range of spacing. Squares, triangles, and circles correspond to island heights of 87 nm, 145 nm, and 125 nm, respectively. Solid black line shows the fit to $1/d^2$ for the black circles; the fit is also good for the blue squares. Note that $T_{BKT}$ depends on island height.

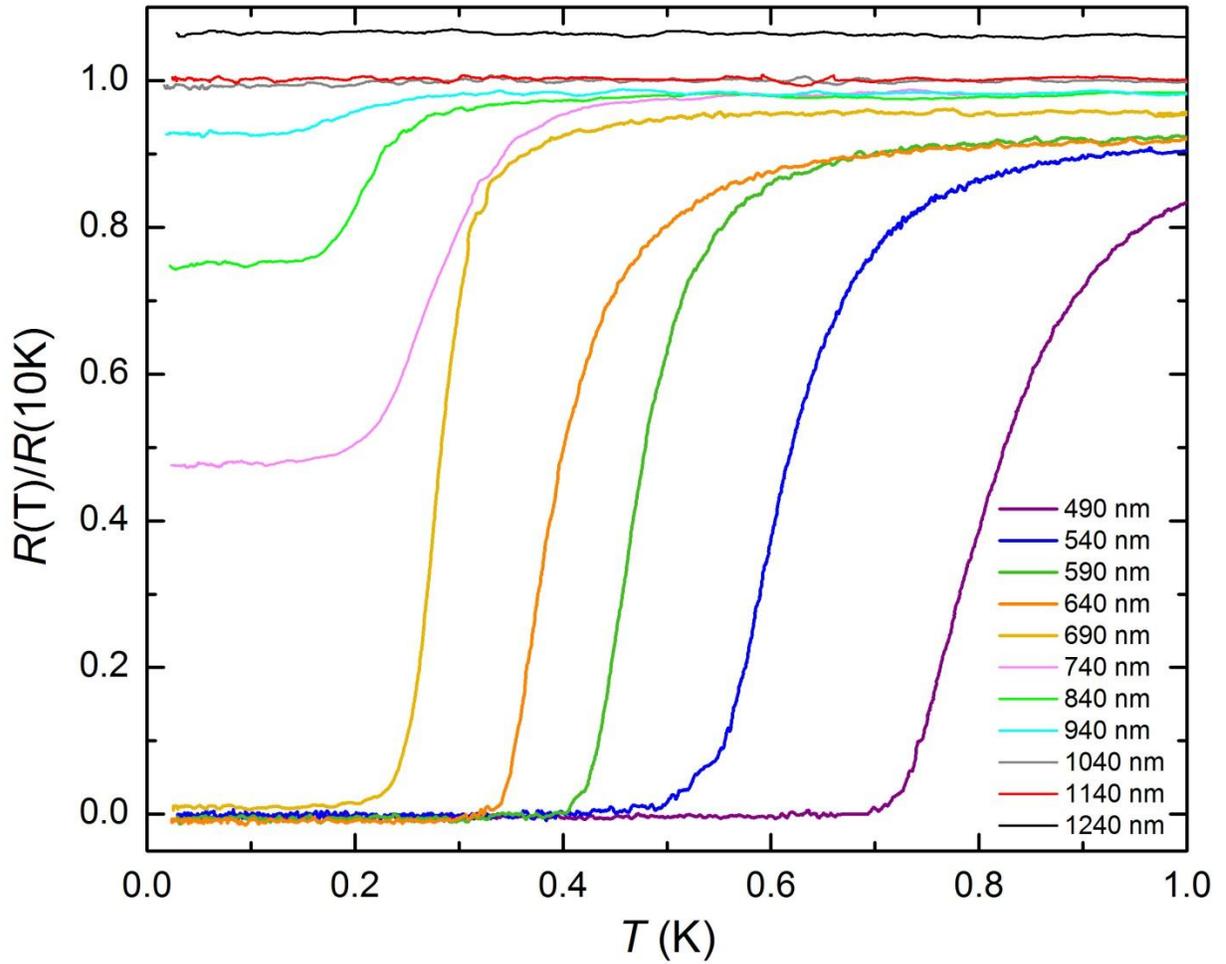

FIG. 4. Normalized resistance as a function of temperature for SNS arrays of widely spaced islands. For spacings exceeding 700 nm, the BKT transition is interrupted by a low-temperature metallic state. The data for $d \leq 690$ nm and for $d \geq 740$ nm come from different substrates, having Nb island heights of 125 nm and 145 nm, respectively.